\documentclass[prd,twocolumn,showpacs,preprintnumbers]{revtex4}
\pdfoutput=1
\usepackage[plainpages=false, colorlinks=true, anchorcolor=blue, linkcolor=blue, citecolor=blue, bookmarks=false]{hyperref}
\usepackage{amsfonts,amsmath,amssymb}
\usepackage{graphicx}
\usepackage{color}
\usepackage{natbib}
\graphicspath{c:/Users/landau/Downloads}
\pdfoutput=1
\begin{document}
%\markboth{Authors' Names}

%%%%%%%% Journals %%%%%%%%%%%%%%
\newcommand{\apjl}{Astrophys. J. Lett.}
\newcommand{\apjs}{Astrophys. J. Suppl. Ser.}
\newcommand{\aap}{Astron. \& Astrophys.}
\newcommand{\aj}{Astron. J.}
\newcommand{\araa}{Ann. Rev. Astron. Astrophys. } %ARA$\&$A}
\newcommand{\mnras}{Mon. Not. R. Astron. Soc.}
\newcommand{\jcap}{JCAP}
\newcommand{\pasj}{PASJ}
\newcommand{\pasa}{Pub. Astro. Soc. Aust.}
\newcommand{\physrep}{Physics Reports.}

%%%%%%%%%%%%%%%%%% TITLE %%%%%%%%%%%%%%%%%%%%%%%%%%%%%%%%%%%% 
\title{GW170817 Falsifies Dark Matter Emulators}
\author{S.  \surname{Boran}$^1$} \altaffiliation{E-mail: borans@itu.edu.tr}
\author{S.  \surname{Desai}$^2$} \altaffiliation{E-mail: shntn05@gmail.com}
\author{E.  O. \surname{Kahya}$^3$} \altaffiliation{E-mail: eokahya@itu.edu.tr}
\author{R.  P. \surname{Woodard}$^4$} \altaffiliation{E-mail: woodard@phys.ufl.edu}

\affiliation{$^{1,3}$Department of Physics, Istanbul Technical University, Maslak 34469 Istanbul, Turkey}
\affiliation{$^{2}$Department of Physics, Indian Institute of Technology, Hyderabad, Telangana-502285, India}
\affiliation{$^{4}$Department of Physics, University of Florida, Gainesville, FL, 32611, USA}

\begin{abstract}

 On August 17, 2017 the LIGO interferometers detected the gravitational wave (GW) signal (GW170817) from the coalescence of binary neutron stars. This signal was also simultaneously seen throughout the electromagnetic (EM) spectrum from radio waves to gamma-rays. We point out that this simultaneous detection of GW and EM signals rules out a class of modified gravity theories, termed ``dark matter emulators,'' which dispense with the need for dark matter by making ordinary matter couple to a different metric from that of GW. We discuss other kinds of modified gravity theories which dispense with the need for dark matter and are still viable. This simultaneous observation also provides the first observational test of Einstein's Weak Equivalence Principle (WEP) between gravitons and photons. We estimate the Shapiro time delay due to the gravitational potential of the total dark matter distribution along the line of sight (complementary to the calculation in~\cite{LVCFermi}) to be about 400 days. Using this estimate for the Shapiro delay and from the time difference of 1.7 seconds between the GW signal and gamma-rays, we can constrain violations of WEP  using the parameterized post-Newtonian (PPN) parameter $\gamma$, and is given by  $|\gamma_{\rm {GW}}-\gamma_{\rm{EM}}|<9.8 \times 10^{-8}$.
\end{abstract}
\maketitle

%%%%%%%%%%%%%%%%%%%%%%%%%%%%%%%%%%%%%%%%%%%%%%%%%%%%%%%%%%%
%%%%%%%%%%%%%%%%%%%%%%%%%%%%%%%%%%%%%%%%%%%%%%%%%%%%%%%%%%%
%%%%%%%%%%%%%%%%%%%%%%%%%%%%%%%%%%%%%%%%%%%%%%%%%%%%%%%%%%%
%\noindent {\it Introduction.---}
\section{Introduction}
The LIGO-Virgo collaboration detected the inspiral and  merger of a binary neutron star (BNS) in the data stream of the LIGO detectors  on August 17, 2017  with very high significance (false-alarm-rate  of less than 1 per  $8 \times 10^4$ years) and combined SNR of about 32,  with the total duration of the detected signal about 100 seconds~\cite{GW170817}. 
A short Gamma-ray Burst (GRB170817A) was detected about 1.7 seconds after this event by the {\it Fermi} Gamma-ray Burst Monitor~\cite{Fermi,LVCFermi,EM}. Soon thereafter, an optical transient was detected  using the SWOPE telescope (designated as SSS17a/AT2017gfo), enabling a precise measurement of its distance and host galaxy~\cite{SWOPE,EM}. 
The position of this transient signal (from the SWOPE observations) is at RA and DEC = $197.45^{\circ}$ and $-23.36^{\circ}$ respectively~\cite{SWOPE}. The host galaxy  of this merger event is NGC 4993, located at a distance of about 40 Mpc~\cite{Freedman}. Subsequently, this signal was also seen in X-rays and radio waves. This is therefore the first GW source for which EM counterparts have been detected.
These observations also provide the first direct evidence that the merger of two neutron stars cause a sGRB and also lead to a kilonova powered by radioactive decay of rapid neutron-capture process (r-process) nuclei elements ejected during the explosion~\cite{EM,kilonova}.

In addition to the above important results, these observations also enable a novel probe of General Relativity (GR) and the equivalence principle for a brand new cosmic messenger, viz.  gravitational waves (GWs). The total time it takes for any carrier  to reach the Earth from the Cosmos is equal to the sum of the distance divided by the (vacuum) speed of light and an additional delay due to the non-zero gravitational potential of the cumulative mass distribution along the line of sight.

The latter delay is known as Shapiro delay~\cite{Shapiro} and has been directly measured for  both a static mass distribution as well  as a moving mass in the Solar system~\cite{Shapiro66,Cassini,Kopeikin}. These solar system measurements have enabled the most precise tests of GR~\cite{Will}. Shapiro delay is also routinely used as an astrophysical tool to measure the masses of neutron stars in binary pulsars~\cite{Taylor94,Demorest,Corongiu12}.

The cumulative Shapiro delay for a cosmic messenger might seem to have only academic interest, both because it can never be measured and because it is much smaller than the vacuum light travel time. However, it does provide an important test of how the various cosmic messengers couple to gravity. The first calculation of this line-of-sight Shapiro delay was done in 1988, following the detection of neutrinos from SN~1987A~\cite{IMB,Kamioka} and the detection of the optical flash about four hours after the neutrino event. It was pointed out, in back-to-back papers~\cite{Longo,Krauss} (see also~\cite{Franson}), that the neutrinos also encountered  a Shapiro delay of about 1-6 months due to the gravitational potential of the intervening matter along the line of sight. We note that this is the only direct evidence  to date that neutrinos are affected by GR and obey WEP to a precision of 0.2-0.5\%.

Recently, there has been a resurgence of interest in the calculations of line-of-sight Shapiro delay following Wei {\it et al.}~\cite{grb} (and citations to it),  who  pointed out how one can constrain the WEP using  simultaneous observations of compact objects throughout the EM spectrum. The violation of WEP in these papers has been quantified in terms of the difference in  PPN parameter  $\Delta \gamma$~\cite{Will}. Consequently, a wide variety of extragalactic as well as galactic astrophysical objects such as Fast Radio Bursts (FRBs)~\cite{Wei,Kaplan,Wu17}, blazars~\cite{blazar}, GRBs~\cite{grb}, and pulsars~\cite{BZhang,YZhang,Desai17} have been used to constrain the WEP. A review of most of these results can be found in Table~1 of Ref.~\cite{Wu17}.
For all these papers, a key ingredient is the accurate calculation of galactic as well as extra-galactic Shapiro delay~\cite{Nusser}. In some of these works, the total gravitational potential of the Laniakea supercluster of galaxies has been considered~\cite{Weijcap}.

A similar test of the WEP for GWs using line of sight Shapiro delay
from a GRB, which simultaneously emits GWs and photons  was first proposed by Sivaram~\cite{Sivaram}. All previous detections by LIGO~\cite{LIGO,GW170104,GW170814,GW15226} were binary black hole mergers, for which no EM counterparts are expected (See, however ~\cite{loebgrb}.) 
However,  the detection of GWs from the first LIGO detection (GW150914) over a frequency range of about 150 Hz within a 0.2 second window, allows us to constrain any frequency-dependent  violations of Shapiro delay for GWs. From GW150914~\cite{LIGO}, one can constrain frequency dependent violations of the WEP for gravitons to within $\mathcal{O} (10^{-9})$~\cite{Kahya16,Gao,maoliu}. Recently, Takahashi~\cite{Takahashi} has pointed out that for a lensing mass of about $\sim 1 M_{\odot}$, gravitational waves in the frequency range of ground-based GW detectors do not experience Shapiro delay, because we are in the geometrical optics regime. However, since galactic masses are $\mathcal{O}(10^{12}~M_{\odot})$, this issue is not of concern to us.

Independently of testing the WEP for different cosmic messengers, the relative Shapiro delay between gravitational waves and photons/neutrinos enables us to test a certain class of modified gravity theories, which dispense with the dark matter paradigm~\cite{Silk} and reproduce Modified Newtonian Dynamics (MOND)~\cite{MOND} like behavior in the non-relativistic limit. Such modified theories of gravity have been dubbed ``Dark Matter Emulators" \cite{Kahya:2007zy}. These models have the property that, in the extreme weak field regime relevant to cosmology, gravitational waves propagate on different geodesics from those followed by photons and neutrinos. Even though the actual model is one of modified gravity, the different geodesics can both be viewed as those of GR, but coupled to different matter sources. The null geodesics of GWs are sourced by only the baryonic matter, however, photons/neutrinos propagate on null geodesics sourced by baryonic matter and the much larger pools of dark matter which would be required if GR were correct. Therefore, the differential Shapiro delay between GWs and photons/neutrinos is due to the gravitational potential of only the dark matter. Some examples of these Dark Matter (DM) emulator theories include Bekenstein's TeVeS theory~\cite{Bekenstein} and Moffat's Scalar-Tensor-Vector gravity theory~\cite{Moffat}. More details on DM emulator theories and some of their predictions made for externally triggered GW searches~\cite{exttrig} during the era of initial and enhanced LIGO can be found in our previous works~\cite{Kahya:2007zy,Kahya08,Desai,Kahya10}. Now that the first simultaneous observation of GWs and photons has occurred, we can carry out our proposed tests.
We note that recently Chesler and Loeb~\cite{Loeb} have pointed out that some relativistic generalizations of MOND show non-linear behavior in the weak field regime, which is inconsistent with observations from GW150914. There are also other severe constraints on TeVeS from binary pulsars~\citep{Friere} and large-scale structure (LSS)~\cite{Dodelson,Reyes}.

The outline of this paper is as follows. We provide a brief summary of  the GW and EM follow-up observations of GW170817 in Sect.~\ref{sec:obs}.  We review the predictions of DM emulator theories and the Shapiro delay calculation in Sect.~\ref{sec:dmemulators}.
Our limits on the violation of WEP are presented in Sect.~\ref{sec:WEP}. We conclude in Sect.~\ref{sec:concl}.

\section{GW170817 observations}
\label{sec:obs}
We provide a brief recap of the GW and EM followup observations of GW170817. More details can be found in the multi-messenger followup paper~\cite{EM}.  
GW170817 was detected by the advanced LIGO and Virgo detectors and the signal was consistent with a binary neutron star coalescence with a merger time at 17th August 2017 12:41:04 UTC~\cite{GW170817}.  The signal lasted for about 100 seconds in the sweet spot of the sensitivity range of the current GW detectors. A corresponding $\gamma$-ray signal was  detected by {\it Fermi} Gamma-Ray Burst Monitor about 1.7 seconds after the merger~\cite{Fermi}. The $\gamma$-ray signal was also confirmed by the Integral satellite~\cite{Integral}.

The first detection of an optical transient was by the One-Meter, Two Hemisphere (1M2H) team, which discovered  a 17th magnitude transient in the $i$-band using the SWOPE telescope~\citep{SWOPE}.  They also pinned down the location of the transient (dubbed SSS17a) to  $\alpha$ (J2000) = 13h09m48.085s and $\delta$ (J2000)=$-23^{\circ}$22'53".343
at a projected distance of 10.6" at the center of NGC 4993 at a distance of about 40 Mpc~\cite{Freedman}. Many other optical teams subsequently confirmed this transient from UV to IR wavelengths. An X-ray counterpart was detected by the Chandra telescope about 9 days after the merger event. Finally, a radio counterpart was detected by the Very Large Array (VLA) about 16 days after the merger event.

\section{Shapiro Delay calculation}
\label{sec:dmemulators}

We define a {\it Dark Matter Emulator} as any modified gravity theory for
which~\cite{Kahya:2007zy}:
\begin{enumerate}
\item{Ordinary matter couples to the metric $\tilde{g}_{\mu\nu}$ ($\tilde{g}$ denotes the ``disformally transformed metric'')
that would be produced by general relativity with dark matter; and}
\item{Gravitational waves couple to the metric $g_{\mu\nu}$ produced by general relativity without dark matter.}
\end{enumerate}

It is important to understand that dark matter emulators constitute a special 
class of modified gravity theories which attempt to dispense with dark matter. Many 
modifications of gravity do not fall within this class, including Milgrom's bi-metric
formulation of MOND~\cite{Milgrom:2009gv} and nonlocal MOND 
\cite{Deffayet:2011sk,Deffayet:2014lba,Kim:2016nnd}. Nor does it include hybrid theories 
such as dipolar dark matter~\cite{dipolar} or superfluid dark 
matter~\cite{superfluid,lasha}. Generalized Einstein-Aether theories 
\cite{Jacobson:2000xp} can be considered dark matter emulators or not, depending upon 
how one choses the vector kinetic term \cite{Lim:2004js,Zlosnik:2006zu}.

When a neutron star merger occurs they emit GWs and photons simultaneously. If physics is described by a DM emulator model then GWs will arrive earlier compared to photons (That they arrive earlier derives from the extra potential due to dark matter. It is also required to avoid the emission of gravitational Cherenkov radiation~\cite{Caves,Moore}.) The additional Shapiro delay in the arrival times of photons would be only due to the dark matter needed if GR was correct. Therefore, one has to include all the contributions coming from the galaxies along the line of sight from NGC 4993 to the Earth.

%\begin{figure}[htbp]
%\centering
%\includegraphics[height=5.0cm]{distances}
%\caption{The distance of the galaxies to the cylindrical line of sight towards NGC 4993}
%\label{distance}
%\end{figure}

\begin{widetext}

\begin{figure}[!tbp]
\centering
\begin{minipage}[b]{0.48\textwidth}
\includegraphics[width=\textwidth]{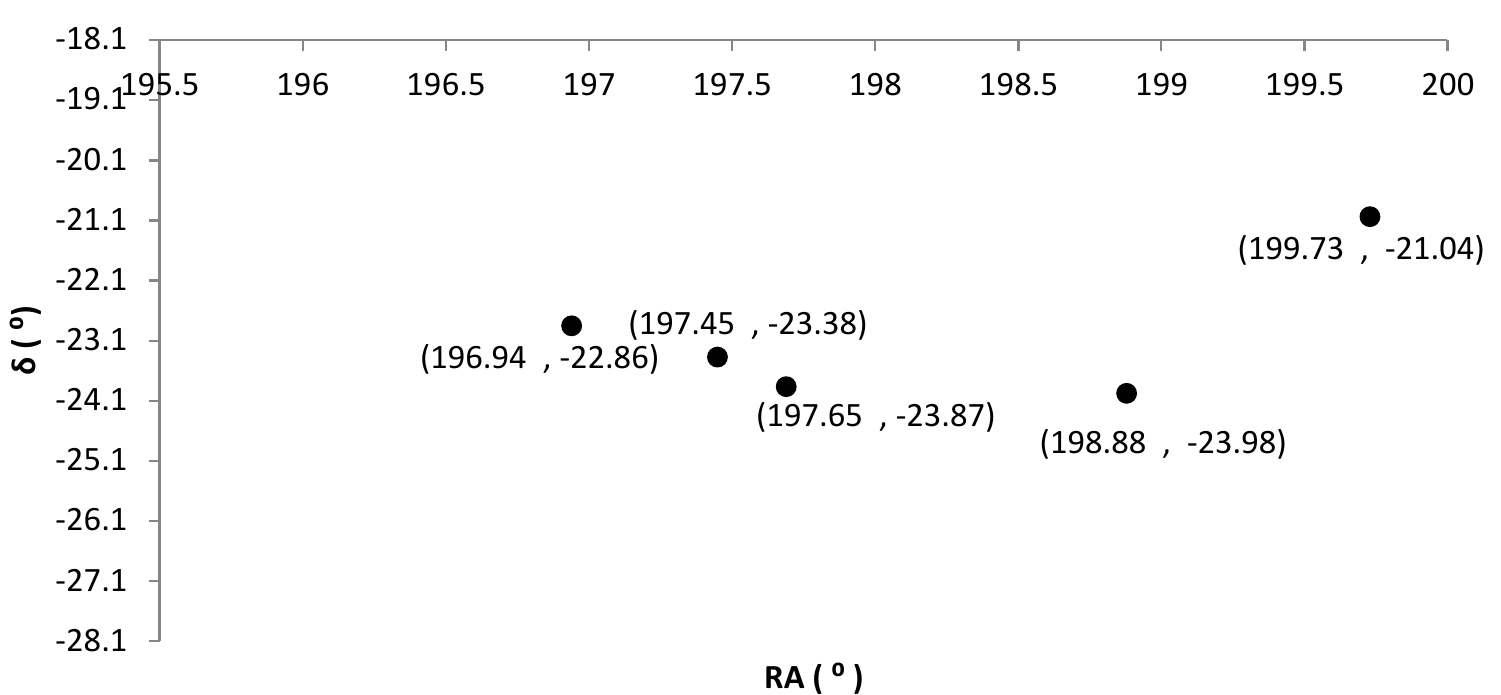}
\caption{The angular locations of galaxies which affect the Shapiro delay of any cosmic messenger coming from NGC 4993}
\label{deg}
\end{minipage}
\hfill
\begin{minipage}[b]{0.5\textwidth}
\includegraphics[width=\textwidth]{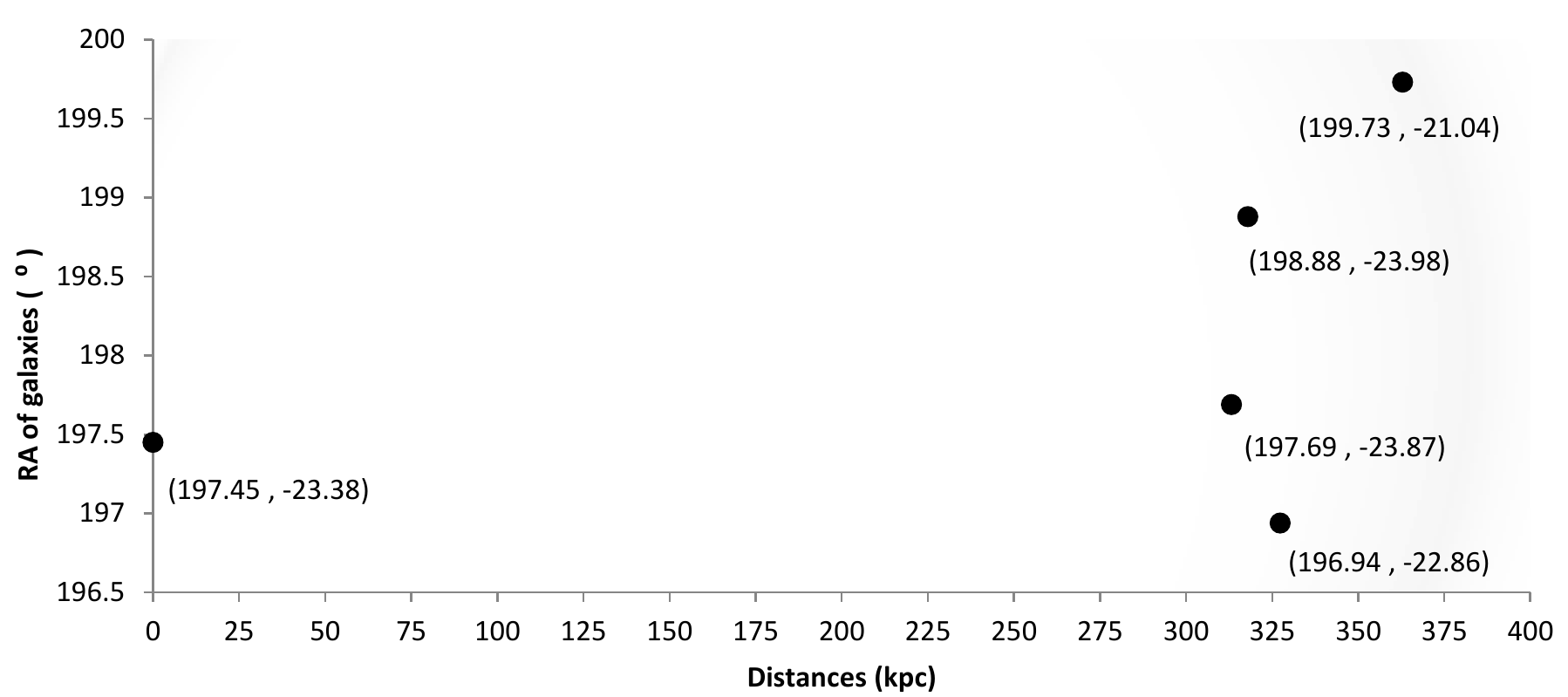}
\caption{The distance of the galaxies along the cylindrical line of sight towards NGC 4993}
\label{distance}
\end{minipage}
\end{figure}

%\begin{widetext}
The main properties of these galaxies are listed in the following Table~\ref{fivegalaxies}. 
\begin{table}[htbp]
\begin{center}
\caption{The properties of the galaxies shown in  Fig.~\ref{distance}}
\begin{ruledtabular}
\begin{tabular*}{1.5\textwidth}{lllllllllll} \\

\textbf{Name} &  \textbf{RA($^\circ$)}& \textbf{$\delta$ ($^\circ$)}&   \textbf{Bmag} & \textbf{e\_Bmag }& 
	
\textbf{PA($^\circ$)} &	\textbf{BMAG} &	\textbf{e\_BMAG }& \textbf{Dist (Mpc)}& \textbf{e\_Dist (Mpc)} \\[0.5ex]

\hline\\

NGC 5068 & 199.73 & -21.04 &	10.20 & 0.30 &
& -18.90 & 0.31 & 6.60 & 1.45\\[2ex]

NGC 5042 & 198.88 & -23.98 & 11.81 &	0.30&
19.0 & -18.70 & 0.31 & 12.65 &	2.53\\[2ex]

ESO 508-011 & 196.94 & -22.86 & 12.88 & 0.30 &
95.0 &-19.26 & 0.31 & 26.79 & 5.36 \\[2ex]

NGC 4993 & 197.45 & -23.38 & 12.87 & 0.19 & 	
173.2 & -20.20 & 0.20 & 33.81 &	5.07\\[2ex]

ESO 508-024 & 197.69 & -23.87&	 12.64& 0.30 &	
72.0 & -19.98 & 0.31 & 33.42 & 5.01\\

\end{tabular*}\label{fivegalaxies}
\end{ruledtabular}
\end{center}
\end{table}

 The columns in Table~\ref{fivegalaxies} are as follows. Bmag: Apparent blue magnitude, 
e\_Bmag (mag): Error in apparent blue magnitude,
PA: Position angle of the galaxy (degrees from north through east),
BMAG: Absolute blue magnitude,
e\_BMAG (MAG): Error in absolute blue magnitude,
Dist (Mpc): Distance (Mpc) and
e\_Dist (Mpc): Error in distance (Mpc).
\end{widetext}

If we consider a cylindrical line of sight, whose radius is 400 kpc from the source to us, then we find using the GLADE catalog (available online at \url{http://aquarius.elte.hu/glade/}), that there are five galaxies with overlaps inside this cylindrical tube~\citep{GLADE}. That means  that the photons and the GWs will be affected by these galaxies. If we look at Fig.~\ref{distance}, the closest galaxy to the cylindrical line of sight is 300 kpc away and therefore the total effect of all 5 galaxies inside the tube is small compared to a single galaxy.

At distances as large as 30 Mpc, Shapiro delay will display logarithmic behavior, if one treats the source as point-like and the metric to be Schwarzschild and is given by:
\begin{equation}
\Delta t_{\rm{shapiro}}=(1+\gamma) \frac{GM}{c^3} \ln\left(\frac{d}{b}\right)\; ,
\end{equation}
where $\gamma$ is the PPN parameter, $b$ is the impact parameter, and $d$ is the distance to the source. For  $M_{\rm{MW}} =10^{12} M_\odot$, $d=400$~kpc,  
$b=8$~kpc, and $\gamma=1$ (assuming GR is correct), this equation gives  $\Delta t^{\rm {MW}}_{\rm{shapiro}} \sim 445$ days~\cite{Kahya08,Desai,Kahya10}. 

Let us assume that the dark matter emulator models mimic cored isothermal profile, and also use recent mass estimate for Milky Way~\cite{Gibbons2014} $M_{\rm{MW}} =5.6 \pm 1.2 \times 10^{11}  M_\odot$. It turns out that the characteristic density for isothermal halo model $\rho_0 = 3.25 \, \rm{GeV/cm^3}$ for a cutoff radius of 200 kpc and core radius of 2 kpc~\cite{Binney2016}. Using these values, one obtains $115\pm 25$ days for the time delay for a source located at 200 kpc. Since the source is now located at a distance of 40 Mpc and that would give a value of $305\pm 65$ days just due to the Milky Way (MW). If we take the contribution due to NGC 4993 of order 100 days similar to MW we get a total time delay as $400\pm 90$ days. The exact number will also depend on the location of the source in NGC4993 galaxy.  Finally, the additional galaxies in between the source and us will have negligible effects because all the galaxies along the cylindrical line of sight are located at positions more than 300 kpc. Therefore based our conservative estimate, we estimate the total Shapiro delay due to the dark matter component of the order of  400 days.

In principle, to obtain a more robust estimate on WEP violation,  the Shapiro delay due to the baryonic matter needs to  be calculated~\cite{Desai15}, but since the total baryonic mass is negligible compared to the total dark contribution, we do not include its effects. However, the baryonic contribution is not needed for testing DM emulator theories.

The precision of this calculation is not important, only the order of magnitude. Because GR predicts coincident arrival times for photons and gravitational radiation, whereas DM emulators predict delays of over a year, the simultaneous optical detection of GW170817 immediately and decisively falsifies DM emulator models.

We also note that an independent estimate of the Shapiro delay was carried in the joint GW-gamma ray observational paper~\cite{LVCFermi}. In that work, they considered the contribution of the Milky way (for which a Keplerian potential with a mass of $2.5\times 10^{11} M_{\odot}$ was assumed) outside a sphere of 100 kpc. 

\section{Constraints on WEP}
\label{sec:WEP}
 Once the Shapiro delay for a given mass distribution is calculated along a line of sight to GW170817, if gravitational waves and photons arrive from the same source  within a time interval ($\Delta t$), after traversing 40 Mpc, one can constrain the violations of WEP in terms of the PPN parameter $\Delta \gamma \;=\; | \gamma_{\rm{GW}} -\gamma_{\rm{EM}} |$ and the calculated Shapiro delay $\Delta t_{\rm{shapiro}}$~\cite{Wei}:
\begin{equation}
\Delta\gamma \leq 2 \; \frac{\Delta t}{\Delta t_{\rm {shapiro}}}\;.
\end{equation}
If we consider the $\Delta t$ to be  time interval between the GRB arrival time (detected by {\it Fermi}) and merger time detected by the LIGO-Virgo detectors and from our calculated value of $\Delta t_{\rm{shapiro}}$  we obtain $\Delta t=1.7$ secs. 
Using this value of $\Delta t$, we get $\Delta \gamma <9.8 \times 10^{-8}$. We note that this limit is more stringent than that obtained in Ref.~\citep{LVCFermi}, which obtained $\Delta \gamma \sim \mathcal{O}(10^{-6}-10^{-7})$,  because in the latter a more conservative estimate of the Shapiro delay has been made.

\section{Conclusions}
\label{sec:concl}
The LIGO-Virgo interferometers  detected the coalescence of a binary neutron star candidate on 17th August 2017 and this GW event has been dubbed as GW170817A. This is the first GW source for which EM counterparts were also detected throughout the spectrum ranging  from $\gamma$-rays (about 1.7 seconds later) to optical (less than  11 hours after), X-rays (9 days later), and radio (16 days later)  after the GW detection~\cite{GW170817,LVCFermi,EM}. These multi-wavelength observations have confirmed the basic picture that  binary neutron star mergers give rise to short GRBs and a kilonova/macronova  caused by r-process nucleosynthesis~\cite{EM,kilonova}. 

Following our previous works~\cite{Desai,Kahya08,Kahya10,Desai15,Kahya16,Desai17}, we calculated the line of sight Shapiro delay from the total dark potential   towards  GW170817 to be about 400 days. This calculation is complementary to a similar estimate done in Ref~\cite{LVCFermi}, which considered the Milky way contribution and assumed a Keplerian potential for the same.
The observations of EM counterparts also allow us to test WEP for photons. From the difference in the arrival times between the $\gamma$-rays and GWs, we point out that gravitons propagate on the same null geodesics, thus obeying WEP. The accuracy of  WEP can be quantified using the difference in PPN $\gamma$ parameters between the GWs and photons and is given by   $|\gamma_{\rm{GW}} - \gamma_{\rm{EM}}|< 9.8 \times 10^{-8}$. 

We also point out that these observations rule out a whole class of modified theories of gravity designed to 
dispense with the need for dark matter, called dark matter emulators. Examples include Bekenstein's TeVeS
theory~\cite{Bekenstein} and Moffat's Scalar-Tensor-Vector gravity theory~\cite{Moffat}. In dark matter emulators
weak gravitational radiation couples to the usual metric which does not carry the extra force needed to compensate
for the absence of dark matter, while normal matter couples to a metric involving additional fields which carry 
the extra force. If these dark matter emulator models were correct, photons from GW170817 would have arrived about 
400  days after the GWs due to the extra Shapiro delay they would experience.

It is important to understand that GW170817 does {\it not} falsify all modified gravity 
models which dispense with dark matter. What it does instead is to place an important 
constraint on how any such model must be constructed. This constraint is just that 
linearized GWs must, with very high precision, couple to the same metric that 
ordinary matter does. Examples of models which meet this requirement are Milgrom's 
bi-metric formulation of MOND~\cite{Milgrom:2009gv} and nonlocal MOND 
\cite{Deffayet:2011sk,Deffayet:2014lba,Kim:2016nnd}. Although bi-MOND involves two 
metrics, the same one which couples to ordinary matter also carries normal gravity waves 
\cite{Milgrom:2013iea}. In nonlocal MOND there is only one metric and gravitational 
radiation is not changed at all because the source of the nonlocal modifications is 
proportional to the Ricci tensor. Our no-go result does not apply to hybrid models 
which replicate MOND phenomenology such as superfluid dark matter~\cite{superfluid} or 
dipolar dark matter~\cite{dipolar}. Nor does it apply to certain types of Einstein-Aether 
theories \cite{Sanders:2011wa,Blanchet:2011wv} whose vector kinetic terms are properly
chosen.

{\bf Note Added:} After this work was submitted  to arXiv, we found three other papers submitted concurrently with similar conclusions~\cite{Meszaros17,Fan,Murase}. Wei et al~\cite{Meszaros17}  considered the Milky way potential and obtained $\Delta \gamma < 5.9 \times 10^{-8}$. Using the potential of the VIRGO cluster, they obtain $\Delta \gamma < 9.1 \times 10^{-11}$. Wang et al~\cite{Fan} considered the potential of the Milky Way and also the potential fluctuations from the large scale structure and obtain $\Delta \gamma < 3.4\times 10^{-9}$. Wang et al have also independently pointed out about the falsification of dark matter emulators. In Ref~\cite{Murase}, two different potentials for the Milky Way were assumed and the estimated limits are $\Delta \gamma<7.4\times 10^{-8}$ and $\Delta \gamma<8.1\times 10^{-7}$.

%%%%%%%%%%%%%%%%%%%%%%%%%%%%%%%%%%%%%%%%%%%%%%%%%%%%%%%%%%%
%%%%%%%%%%%%%%%%%%%%%%%%%%%%%%%%%%%%%%%%%%%%%%%%%%%%%%%%%%%
%%%%%%%%%%%%%%%%%%%%%%%%%%%%%%%%%%%%%%%%%%%%%%%%%%%%%%%%%%%

\begin{acknowledgements}
E.O.K. acknowledges support from TUBA-GEBIP 2016, the Young Scientists Award Program. R.P.W. acknowledges support from NSF grant PHY-1506513, and from the Institute for Fundamental Theory at the University of Florida.
\end{acknowledgements}

% LocalWords:  PSR
\bibliography{dmemulator}
\end{document}